\documentclass[12pt,a4paper]{article}

\usepackage[british]{babel}

\usepackage[a4paper,top=2cm,bottom=2cm,left=2.5cm,right=2.5cm,marginparwidth=1.75cm]{geometry}

\usepackage{amsmath}
\usepackage{graphicx}
\usepackage[colorlinks=true, allcolors=blue]{hyperref}
\usepackage{hyperref}
\usepackage{orcidlink}
\usepackage[title]{appendix}
\usepackage{mathrsfs}
\usepackage{amsfonts}
\usepackage{booktabs} 
\usepackage{caption}  
\usepackage{threeparttable} 
\usepackage{algorithm}
\usepackage{algorithmicx}
\usepackage{algpseudocode}
\usepackage{listings}
\usepackage{enumitem}
\usepackage{chngcntr}
\usepackage{booktabs}
\usepackage{lipsum}
\usepackage{subcaption}
\usepackage{authblk}
\usepackage[T1]{fontenc}    
\usepackage{csquotes}       
\usepackage{diagbox}

\usepackage{cite}

\usepackage{setspace}
\usepackage{titlesec}
\titleformat{\section} 
  {\normalfont\Large\bfseries}{\thesection.}{1em}{}
  
\usepackage{float}   
\usepackage{caption} 


\title{\textbf{Vacancy-induced Modification of Electronic Band Structure of LiBO$_{2}$ Material as Cathode Surface Coating of Lithium-ion Batteries}}
\author[1,2,*,**]{Ha M. Nguyen}
\author[1,3,**]{Carson D. Ziemke}  
\author[2]{Narendirakumar Narayanan}
\author[5]{Sebastian Amaya-Roncanci}
\author[2]{John Gahl}
\author[1,4]{Yangchuan Xing}
\author[1,2,3]{Thomas W. Heitmann}
\author[1,3,*]{Carlos Wexler}
\affil[1]{\small Materials Sciences and Engineering Institute, University of Missouri, Columbia, MO 65201, USA}
\affil[2]{University of Missouri Research Reactor (MURR), University of Missouri, Columbia, MO 65203, USA}
\affil[3]{Department of Physics and Astronomy, University of Missouri, Columbia, MO 65201, USA}
\affil[4]{Department of Chemical and Biomedical Engineering, University of Missouri, Columbia, MO 65201, USA}
\affil[5]{Escuela de Física, Universidad Pedagógica y Tecnológica de Colombia, Tunja, Colombia}
\affil[*]{Corresponding authors: \texttt{hn4gq@missouri.edu, wexlerc@missouri.edu}}
\affil[**]{Authors with equal contributions}

\begin{document}
\maketitle

\begin{abstract}

LiBO$_{2}$ is an electronic insulator and a promising surface coating for stabilizing high-voltage cathodes in lithium-ion batteries. Despite its potential, the functional mechanisms of this coating remain unclear, particularly the transport of lithium ions and electrons through LiBO$_{2}$ in the presence of lattice vacancies. This understanding is critical for the design and development of LiBO$_{2}$-based materials. In our previous work [Ziemke \textit{et al.}, J. Mater. Chem. A, 2025, \textbf{13}, 3146-3162], we used density functional theory (DFT) calculations to investigate the impact of lattice vacancies on Li-ion transport in both tetragonal (t-LBO) and monoclinic (m-LBO) polymorphs of LiBO$_{2}$, revealing that B vacancies in either polymorph enhanced lithium-ion transport. In this study, we expand on these findings by using DFT calculations to examine the effects of lattice vacancies on the electronic properties of both t-LBO and m-LBO polymorphs,focusing on the electronic band structure. Our analysis shows that B vacancies can enhance the electronic insulation of t-LBO while improving the ionic conduction of m-LBO. The combined results of our previous and current works indicate that B vacancy generation in LiBO$_{2}$ may enable t-LBO to function as a promising solid electrolyte and enhance the performance of m-LBO as a conformal cathode coating in lithium-ion batteries. Overall, generating B vacancies, such as through neutron irradiation, would offer a viable strategy to improve the functionality of LiBO$_{2}$ as a promising material for energy storage applications.
 
\end{abstract}

\noindent
\textbf{Keywords}: lithium metaborate, solid electrolytes, cathode coatings, density functional theory, electronic band structure, lattice vacancy, Li-ion batteries.  



\section{Introduction \label{Intro}}

\noindent

Lithium metaborate (LiBO$_{2}$) is an electronic insulator that has drawn increasing research attention due to its multiple technological applications \cite{Islam2012,Jiang2020,Ogorofnikov,Zebarjad}. These include the surface coating of the cathode of liquid electrolyte-based lithium-ion batteries (LIBs) \cite{Gao2021, Guo2023,Ramkumar,Zhang}. The performance and efficiency of LIBs are critically dependent on the properties of the materials used in their construction, particularly those of the cathode and its coating \cite{Guan2020,Tan2020,Kaur2022,Maske2024}. High-voltage cathode materials, which have been developed for LIBs operating at voltages as high as 3 to 5 V, include layered oxides LiMO$_{2}$ (M = Ni$_{x}$Mn$_{1-x-y}$Co$_{y}$ with $0 \leq x, y, z \leq 1$), spinel oxides LiM$_{2}$O$_{4}$ (M = Ni$_{x}$Mn$_{2-x}$), and polyanion compounds like LiMXO$_{4}$ (M = Fe, Ni, Co, and Mn; X = B, P, Si, W, Mo, etc.) \cite{Kaur2022,Ghosh2022}. However, such a high-voltage window can cause serious stability problems for the liquid electrolyte, the cathode itself, and hence for the overall performance of the LIB \cite{Kaur2022,Ghosh2022,Hou2023}.  For example, metal dendrite formation on the cathode surface, cathode surface degradation, decomposition of unstable components of the liquid electrolyte, and side reactions between the cathode and liquid electrolyte are some of the problems \cite{Kaur2022,Ghosh2022,Hou2023,Li2023}. 

These problems have challenged the research community to seek effective strategies to mitigate them. To date, two of the most-studied approaches for the materials chemistry of LIBs are: (i) modification of the liquid electrolyte by means of additives in order to improve the stability of the solid electrolyte interphase (SEI) between the liquid electrolyte and the surface of each of the electrodes, and (ii) modification of the cathode itself by doping one or more structurally-stabilizing element(s) into the lattice structure of the cathode, or by introducing one or more physical barrier layer(s) on the cathode surface to passivate unwanted side reactions \cite{Kaur2022,Ghosh2022,Hou2023}. This passivation of the cathode has been known as the cathode coating of LIBs, which has been studied for more than a decade with numerous coating materials \cite{Guan2020,Tan2020,Kaur2022,Maske2024}.  Although traditional coating materials, such as Li-free electronic insulators (e.g., Al$_{2}$O$_{3}$, AlF$_{3}$, MgO, ZrO$_{2}$, SiO$_{2}$), have demonstrated their ability to enhance the overall performance of LIBs, their limited Li-ion conductivity and solubility result in increased overpotentials \cite{Xu2015}. This means that more energy is required for the coated electrode compared to that of its uncoated counterpart to achieve the same electrochemical performance. These limitations have driven researchers to explore alternative coating materials, including lithium-containing compounds such as LiAlO$_{2}$ \cite{Park2014, Wang2024} and LiBO$_{2}$ \cite{Gao2021, Guo2023,Ramkumar,Zhang}, whose electronic and Li-ion conduction models resemble those of SEIs of LIBs \cite{Xu2015}.

LiBO$_{2}$, in particular, has recently been found to have significant potential to stabilize cathode materials, especially some layered oxides LiMO$_{2}$ (M = Ni$_{x}$Mn$_{1-x-y}$Co$_{y}$ with $0 \leq x, y, z \leq 1$), resulting in their increased overall electrochemical performance. Zhang \textit{et al.} \cite{Zhang} synthesized LiBO$_{2}$-coated LiNi$_{0.6}$Co$_{0.2}$Mn$_{0.2}$O$_{2}$ for use in polyetheracrylate-based solid-state batteries, reporting a capacity retention of 84.3\% after 150 cycles at 0.5 C. Wang \textit{et al.} \cite{Wang2015} found that LiBO$_{2}$ coating on LiNi$_{0.5}$Co$_{0.2}$Mn$_{0.3}$O$_{2}$ significantly improved its performance, achieving a retention of 90.1\% after 100 cycles at 1 C (3.0-4.6 V), compared to 72.3\% for the unmodified material. Du \textit{et al.} \cite{Du2019} applied a wet chemical method to coat LiBO$_{2}$ on LiNi$_{0.8}$Co$_{0.1}$Mn$_{0.1}$O$_{2}$ (NCM811), with the modified sample delivering a capacity of 157.7 mAh g$^{-1}$ and a retention of 82.1\% after 100 cycles at 1 C and 4.5 V, outperforming the pristine material, which only had a capacity of 96.0 mAh g$^{-1}$ and retention of 50.8\%. Lim \textit{et al.} \cite{Lim2014} employed a solution method to coat LiNi$_{0.8}$Co$_{0.15}$Al$_{0.05}$O$_{2}$ with LiBO$_{2}$, resulting in an initial capacity of 173 mAh g$^{-1}$ and a retention of 94.2\% after 100 cycles at 1 C and 55 °C (3-4.3 V), which was superior to the unmodified sample’s 75.3\% retention.

Interestingly, LiBO$_{2}$ material coating on the surface of the cathode and doping boron into its crystal lattice have been reported to synergistically enhance the overall performance of LIBs even further \cite{Gao2021,Guo2023}. For instance, Gao \textit{et al.} \cite{Gao2021} reported that doping the cathodes with boron and coating them with LiBO$_{2}$ simultaneously boosted the electrochemical performance of the LiNi$_{0.6}$Co$_{0.1}$Mn$_{0.3}$O$_{2}$ (NCM613) cathode material operating at high charging/discharging rates, high temperatures, and high voltage cutoffs. Specifically, they revealed that their optimized NCM613 cathode coated with LiBO$_{2}$ and doped with boron has shown its outstanding cycling stability at room temperature, retaining 94.8\% of its capacity at 1 C after 100 cycles and 70.7\% at 5 C after 1000 cycles (2.8-4.5 V). At 45$^{\circ}$C and 1 C, this modified NCM163 cathode also exhibited an initial discharge capacity of 195.8 mAh g$^{-1}$ and maintained 88.0\% of its capacity after 100 cycles, significantly outperforming the pristine NCM163 material, which only retained 67.3\%. Moreover, the modified NCM163 material showed a higher lithium diffusion coefficient compared to its pristine NCM613 counterpart. Therefore, this work suggested an alternative approach to modify the cathode materials of LIBs using the B element. This suggestion is in line with published works \cite{Guzman-Gonzalez2023,Liua2015} demonstrating that the recently-developed family of boron-containing liquid electrolytes, which resulted in a favorable SEI, stabilizing and enhancing the performance of LIBs cathode materials.

Because of its aforementioned potential as an effective cathode coating, recent basic research focusing on ionic and electronic properties of LiBO$_{2}$ has been increasingly revived in order to provide insights into the design and development of more effective LiBO$_{2}$-based cathode coatings. Hirose \textit{et al.} \cite{Hirose2019} conducted an experimental study on Li-ion transport in high-density polycrystalline LiBO$_{2}$ polymorphs synthesized under high pressures, the tetragonal LiBO$_{2}$ (t-LBO) and the monoclinic LiBO$_{2}$ (m-LBO). It is noted that while the t-LBO is only stable at high temperatures and/or high pressures, the m-LBO is stable at ambient temperature and atmospheric pressure. Using electrochemical impedance spectroscopy (EIS) of their samples between 450 and 520 K, they found that the intra-grain and total conductivities of the t-LBO  with a three-dimensional Li-Li network were between 10$^{-6}$ and 10$^{-5}$ S cm$^{-1}$, with total conductivity ca.\ 10$^{-6}$ S cm$^{-1}$. These values were consistently higher than those observed for the m-LBO with a two-dimensional network across the entire temperature range. Islam \textit{et al.} \cite{Islam2011} conducted density functional theory (DFT) calculations for the formation and diffusion of Li defects in m-LBO crystal and the effect of Li defects on the electronic density of states (DOS). They found that the defect modified the DOS of the perfect crystal with the presence of defect levels in the band gap. They also found that there are numerous possible pathways of Li$^{+}$ diffusion in the m-LBO, including those along the xy plane.

In our previous work \cite{Ziemke1stDFT_LiBO2}, we systematically examined the formation of Li, B, and O vacancies in both m-LBO and t-LBO and their effects on Li-ion transport using DFT calculations. We found that the formation energy of lattice vacancies increases from Li to O, then to B. For Li-ion transport, O vacancies lower the migration energy barrier ($E_{\text{m}}$) in m-LBO but increase it in t-LBO. In contrast, B vacancies reduce $E_{\text{m}}$ in both m-LBO and t-LBO, enhancing diffusivity and ionic conductivity. Thus, our previous work suggested that creating B vacancies using such an approach as neutron irradiation \cite{{Was2007}} may improve ionic conductivity in the LiBO$_{2}$ material. Basalaev \textit{et al.} \cite{Basalaev2019} have studied the electronic band structures of both m-LBO and t-LBO using DFT calculations and shown that the values of the band gap are 10.4 eV and 7.6 eV for t-LBO and m-LBO, respectively. Nevertheless, how the lattice vacancies affect the electronic band structures of these polymorphs is still an open question. Answering this question is of paramount importance for better designing and engineering conformal LiBO$_{2}$-based surface coatings of the cathode \cite{Guan2020,Tan2020,Kaur2022,Maske2024}. This is because a conformal surface coating is supposed to simultaneously hold multiple conflicting protective mechanisms of its functionalities, including ionic and electronic transport properties, which are still poorly understood \cite{Guan2020,Tan2020,Kaur2022,Maske2024,Xu2015}.

In this work we address the aforementioned question, aiming at the effects of lattice vacancies on the electronic band structures of both m-LBO and t-LBO (see Figure \ref{Figure1}) using DFT calculations, the setup of which is presented in Section \ref{Methods}. Section \ref{r&d} presents and discusses the outcome of our DFT calculations, where we show that boron vacancies have significant potential to enhance the properties of LiBO$_2$ as both a solid state electrolyte and a conformal cathode coating.

\section{Computational Methods\label{Methods}}

DFT calculations were performed to investigate the electronic band structures of both pristine and defective crystals of the m-LBO (space group P2$_{1}$/c)  and t-LBO (space group I$\bar{4}$2d) polymorphs of LiBO$_{2}$ material. These calculations were carried out using the WIEN2K code \cite{Schwarz2003}, which is based on the full-potential linearized augmented plane wave (FP-LAPW) and local orbital basis set method to solve Kohn–Sham equations to calculate the electronic band structures of the LiBO$_{2}$ solid crystals \cite{Blaha1990}.

The Perdew-Burke-Ernzerhof (PBE) form of the Generalized Gradient Approximation (GGA) was employed for the exchange-correlation energy functional \cite{PBE1996}. The core and valence electrons were treated self-consistently within the all-electron framework. For both polymorphs, a convergence criterion of 0.0001 Ry for the total energy was applied.

A Monkhorst-Pack $k$-point mesh of $5\times5\times5$ $k$-points was adopted for the Brillouin zone sampling and the Brillouin zone  integration is carried out using the modified tetrahedron method \cite{Bloechl}, and the plane wave cut-off parameter $R_{K\text{max}}$ was set to 7.00, ensuring adequate basis set completeness. The Muffin-tin radii were carefully chosen to avoid overlap and were respectively set to $1.61$, $1.14$, and $1.32$ Bohr for Li, B, and O atoms in m-LBO and were respectively set to $1.66$, $1.30$, and $1.51$ Bohr for Li, B, and O atoms in t-LBO. 

Defective lattice structures were modeled by introducing one lattice vacancy per supercell at either Li or B or O site into the respective supercells for each of the m-LBO and t-LBO polymorphs constructed from the fully optimized pristine (\text{i.e.}, perfect or nondefective) crystal structures, which were subject to full relaxation of atomic positions and lattice parameters until the force acting on each atom was less than $1$ mRy/Bohr) (See Ref. \cite{Ziemke1stDFT_LiBO2}). Afterwards, the defective supercells were also fully relaxed until the aforementioned criteria of total-energy and force convergences were fulfilled. In this work, we have mainly focused on $1\times1\times1$ supercells so that we could attain highly concentrated lattice vacancies. Our goal is to benchmark an extreme case (about $8\times10^{23} \text{ vacancies/cm}^{3}$, i.e., one vacancy per unit cell), as discussed in our previous work \cite{Ziemke1stDFT_LiBO2}. Although quite high from a materials engineering perspective, this would be relevant in a scenario where LiBO$_{2}$ coating of the cathode in LIBs that are subject to intense or prolonged neutron irradiation \cite{Ziemke1stDFT_LiBO2} such as LIBs operating in extreme environments such as nuclear reactors or deep-space exploration spacecrafts \cite{Qiu2015,Tan2016,Li2018,Pathak2023}.

Experimentally, neutrons, in particular thermal neutrons, can generate either Li or B vacancies in compounds containing these elements such as LiBO$_{2}$ by transmuting either the $^{10}$B (19.9\% of B) or $^{6}$Li (7.5\% of Li) isotopes at their respective lattice sites via the  $^{10}_{5}\text{B} + ^{1}_{0}\text{n (0.025 eV)} \rightarrow ^{4}_{2}\text{He (1.47 MeV)} + ^{7}_{3}\text{Li (0.84 MeV)}$ + $\gamma$ (0.48 MeV) or the  $^{6}_{3}$Li + $^{1}_{0}$n (0.025 eV)  $\rightarrow$ $^{3}_{1}$H (2.73 MeV) + $^{4}_{2}$He (2.05 MeV) + 4.78 MeV nuclear reactions \cite{John2020,John2023}. As observed in the outcomes of our experimental characterization of the m-LBO surface irradiated with thermal neutrons at various neutron doses (not shown here, will be reported elsewhere \cite{HaLiBO2experimental}), the concentration of lattice vacancies was found to be dose-dependent. Accordingly, for sufficiently high neutron doses, the vacancies were no longer isolated, but aggregated into clusters and mutually interacting. Thus, our $1\times1\times1$ supercells with one B vacancy per unit cell, which contains four B atoms, could model the case that all of $^{10}$B sites  are completely transmuted with intense and/or prolonged exposure of the material to thermal neutrons, e.g., in the extreme environments mentioned above. 

In principle, one could investigate computationally the generalized supercells of defective crystals of the form Li$_{1-x}$B$_{1-y}$O$_{2-z}$ with $0 \leq x,y < 1$ and  $0 \leq z < 2$. In this work, we only focus on a limited number of cases for each of the two polymorphs, namely, ($x = y = z =0$, \textit{i.e.}, pristine), ($x=0.25$ and $y = z =0$, \textit{i.e.}, Li vacancy), ($y = 0.25$ and $x=z=0$, \textit{i.e.,} B vacancy),
and ($z = 0.125$ and $x=y = 0$, \textit{i.e.,} O vacancy). Other cases may be considered in the future.

\begin{figure}[!ht]
\centering
\includegraphics[width=1.0\linewidth]{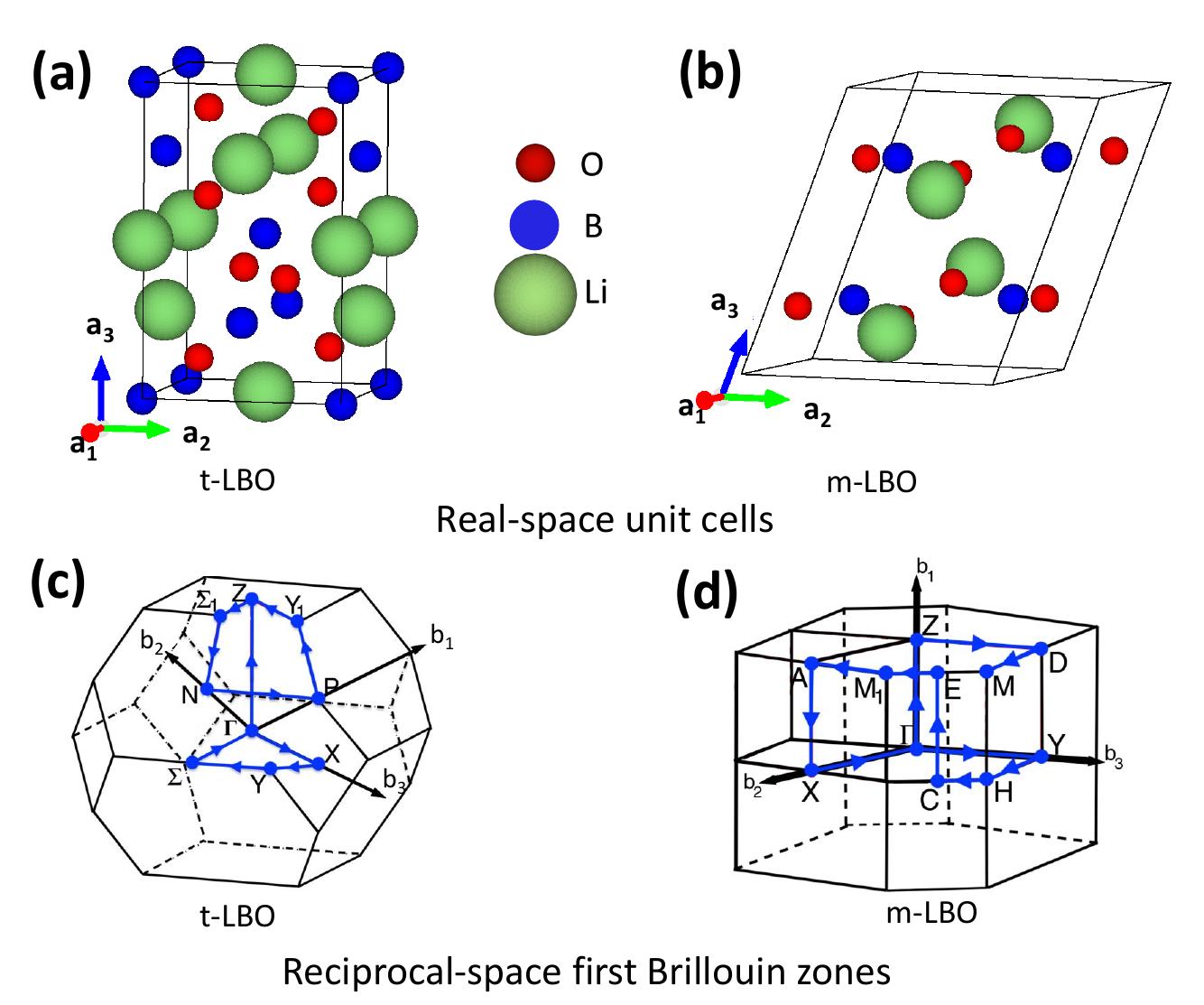}
\caption{\label{Figure1} Structures of LiBO$_{2}$ in real and reciprocal spaces. Unit cells of t-LBO (a) and m-LBO (b) polymorphs of LiBO$_{2}$ crystal. First Brillouin zone of t-LBO (c) and m-LBO (d).}
\end{figure}

The electronic band structures of both pristine and defective crystals were calculated along high-symmetry $k$-paths of the first Brillouin zones \cite{Setyawan} as shown in Figures \ref{Figure1}c and \ref{Figure1}d. The $k$-paths were created using the XCrysDen software \cite{Kokalj}. Specifically, for the t-LBO polymorph of a centered tetragonal lattice structure ($\mathbf{a}_{1}=(a, 0, 0)$, ${\mathbf{a}_{2}}=(0, b, 0)$, and $\mathbf{a}_{3}=(0, 0, c)$, $a{ } ={ } b{ } <{ } c, \alpha = \beta = \gamma$), the $k$-path (see Figure \ref{Figure1}c) is
$\Sigma$ $\rightarrow$ $\Gamma$ $\rightarrow$ X $\rightarrow$ Y $\rightarrow$ $\Sigma$ $\rightarrow$ $\Gamma$ $\rightarrow$ Z $\rightarrow$ $\Sigma_{1}$ $\rightarrow$ N $\rightarrow$ P $\rightarrow$ $\text{Y}_{1}$ $\rightarrow$ Z, of which the fraction coordinates of the $\mathit{k}$-points are $\Sigma (-\eta,\eta,\eta)$, $\Gamma(0,0,0)$, X(0,0,$\frac{1}{2}$), Y($-\zeta$,$\zeta$,$\frac{1}{2}$), Z($\frac{1}{2}$,$\frac{1}{2}$,$-\frac{1}{2}$), $\Sigma_{1}(\eta,1-\eta,-\eta)$, N(0, $\frac{1}{2}$,0), P($\frac{1}{4}$,$\frac{1}{4}$,$\frac{1}{4}$), and Y$_{1}$($\frac{1}{2}$,$\frac{1}{2}$,$\eta$) given the reciprocal lattice vectors being respectively $\mathbf{b}_{1}(\frac{2\pi}{a},0,\frac{2\pi}{a})$, $\mathbf{b}_{2}(\frac{2\pi}{a},\frac{2\pi}{a},0)$, and $\mathbf{b}_{3}(0,\frac{2\pi}{c},\frac{2\pi}{c})$, where $\eta =\frac{1+a^{2}/c^{2}}{4}$ and $\zeta = \frac{a^{2}}{2c^{2}}$ are determined from the corresponding lattice parameters $a$ (note that $a=b$ for a tetragonal lattice) and $c$ of the corresponding pristine or defective tetragonal crystals (see Ref \cite{Ziemke1stDFT_LiBO2}).

In contrast, for the m-LBO polymorph of a monoclinic lattice structure ($\mathbf{a}_{1}=(a, 0, 0)$, ${\mathbf{a}_{2}}=(0, b, 0)$, and $\mathbf{a}_{3}=(0, c\cos\alpha, c\sin\alpha)$, where $a,{ } b{ } < { }c,{}\alpha{ }< 90^{\text{o}}, \beta{ }={ }\gamma{ }=90^{\text{o}}$), the $k$-path (see Figure \ref{Figure1}d) is $\Gamma$ $\rightarrow$ Y $\rightarrow$ H $\rightarrow$ C $\rightarrow$ E $\rightarrow$ M$_{1}$ $\rightarrow$ A $\rightarrow$ X $\rightarrow$ $\Gamma$ $\rightarrow$ Z $\rightarrow$ D $\rightarrow$ M with the fractional coordinates of the high-symmetry $k$-points being $\Gamma (0,0,0)$, $\text{Y} (0,0,\frac{1}{2})$, $\text{H} (0,\eta,1-\nu)$, $\text{C} (0,\frac{1}{2}, \frac{1}{2})$, $\text{E} (\frac{1}{2},\frac{1}{2}, \frac{1}{2})$, $\text{M$_{1}$} (\frac{1}{2},1-\eta,\nu)$, $\text{A} (\frac{1}{2},\frac{1}{2}, 0)$, $\text{X} (0,\frac{1}{2}, 0)$, $\text{Z} (\frac{1}{2},0, 0)$, $\text{D} (\frac{1}{2},0, \frac{1}{2})$, and $\text{M} (\frac{1}{2},\eta,1-\nu)$ given the reciprocal lattice vectors being respectively $\mathbf{b}_{1}(\frac{2\pi}{a},0,0)$, $\mathbf{b}_{2}(0,\frac{2\pi}{b},\frac{2\pi\cot\alpha}{b})$, and $\mathbf{b}_{3}(0,0,\frac{2\pi}{c\sin\alpha})$, where $\eta =\frac{2-(b/c)\cos(\alpha)}{2\sin^{2}(\alpha)}$ and $\nu = \frac{1}{2}-\eta\frac{c}{b}\cos\alpha$ are determined from the corresponding lattice parameters $a$, $b$, $c$, and $\alpha$ of the corresponding pristine or defective monoclinic crystals (see Ref \cite{Ziemke1stDFT_LiBO2}).

\section{Results and Discussions\label{r&d}} 

This section focuses on presenting and discussing the outcome of our DFT calculations of electronic band structures for both t-LBO (Section \ref{t-LBO}) and m-LBO (Section \ref{m-LBO}) polymorphs. 

\subsection{Effects of Lattice Vacancies on Electronic Band Structure of Tetragonal LiBO$_{2}$ \label{t-LBO}}

\begin{figure}[!ht]
\centering
\includegraphics[width=1.0\linewidth]{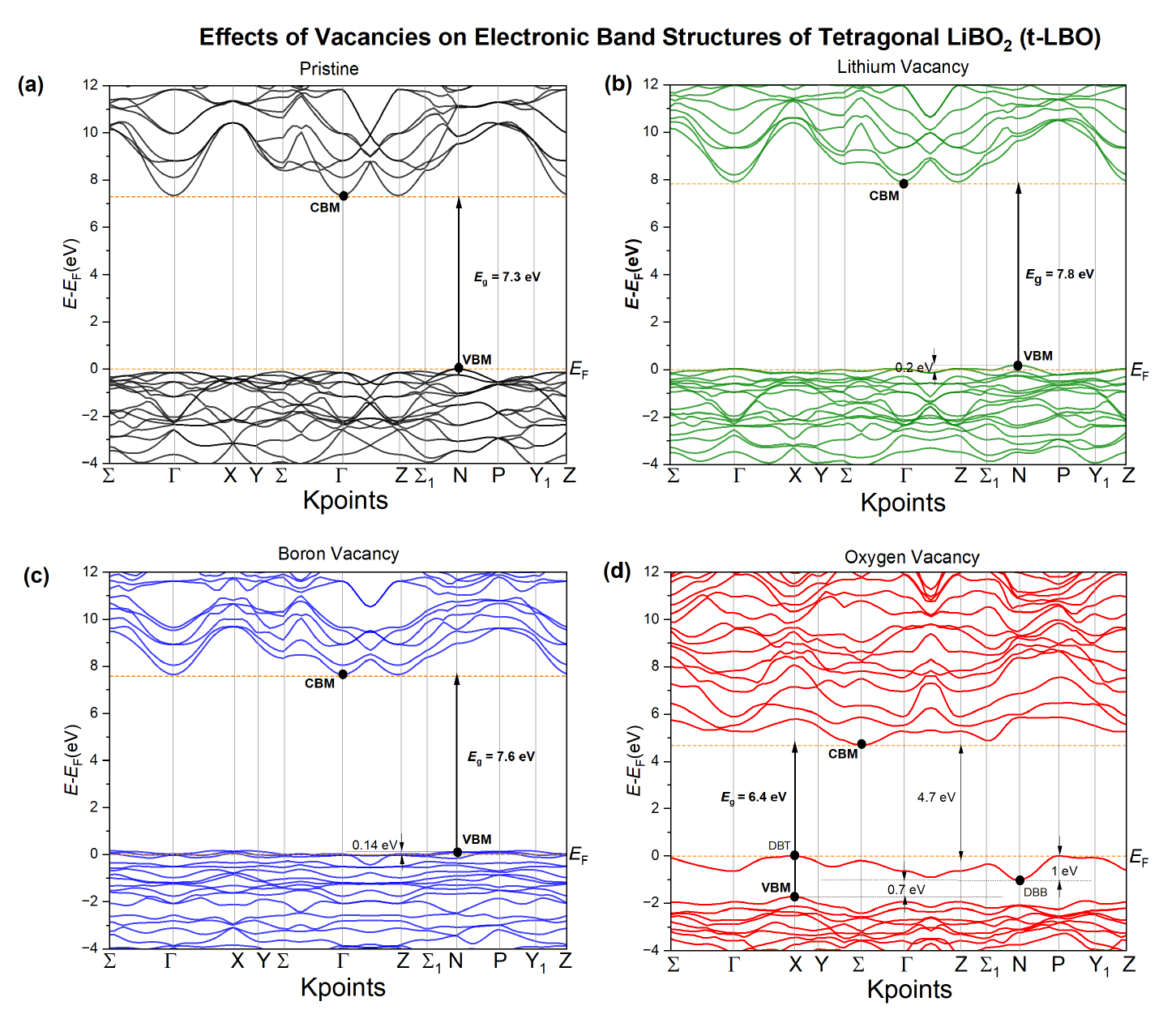}
\caption{\label{Figure2} Calculated electronic band structure of tetragonal LiBO$_{2}$ (t-LBO) for (a) pristine, (b) Li-vacancy, (c) B-vacancy, and (d) O-vacancy supercells. The conduction band minimum, valence band maximum, defect band top, and defect band bottom are abbreviated as CBM, VBM, DBT, and DBB, respectively.}
\end{figure}

Figure \ref{Figure2} shows the electronic band structures calculated for the tetragonal LiBO$_{2}$ polymorph: pristine (Pri t-LBO, Figure \ref{Figure2}a), Li-vacancy (Livac t-LBO, Figure \ref{Figure2}b), B-vacancy (Bvac t-LBO, Figure \ref{Figure2}c), and O-vacancy (Ovac t-LBO, Figure \ref{Figure2}d) supercells.

\textbf{Pristine Tetragonal LiBO$_{2}$} \textbf{(Pri t-LBO)}: Figure \ref{Figure2}a presents the calculated electronic band structure of the Pri t-LBO, which manifests an indirect band-gap electronic band structure with the valence band maximum (VBM) located at the N $k$-point of the Fermi level and the conduction band minimum (CBM) located at the $\Gamma$ $k$-point and separated from the VBM by a band gap of $E_{\text{g}} = 7.3\text{ eV}$, the value that is less than that ($E_{\text{g}}=10.4\text{ eV}$) calculated by Basalaev \textit{et al.} \cite{Basalaev2019} using a computationally demanding hybrid functional method and is attributed to the limitation of a less computationally demanding GGA method used in our current work to underestimate the value of $E_{\text{g}}$. 

It is noted that, (1) the energy level of the top of the valence band at the $\Gamma$ and the Z $k$-points are very close ($\approx 0.2 \text{ eV}$) to the Fermi level, and (2) the energy levels of the bottom of the conduction band at $\Gamma$ and Z $k$-points are also very close ($\approx 0.2 \text{ eV}$). Therefore, the cross-band-gap $\Gamma$-to-$\Gamma$ or Z-to-Z excitations of the electron from the top of the valence band to the bottom of the conduction band might need energies approximately of the order of $E_{\text{g}}=7.3\text{ eV}$ without the need of a phonon absorption to fulfill the law of momentum conservation, suggesting that the Pri t-LBO might behave in reality as a quasi-indirect band gap insulator at elevated temperatures under a stimulation of sufficiently strong electric potentials or an absorption of extreme/deep ultraviolet (EUV/DEV) photons. 

Finally, regarding the electric properties of charge carriers, Figure \ref{Figure2}a shows that the curvatures of the lowest energy dispersion relation $E(k)$ curve of the conduction band at $\Gamma$ and Z $k$-points are quite similar and are both larger than those of the highest $E(k)$ curve of the valence band at these $k$-points, suggesting that the effective mass of an electron at the bottom of the conduction band is less than that of an electronic hole at the top of the valence band, and hence the electron, if excited to the conduction band, would likely be more mobile than the electronic hole that is excited to the valence band. Recall that the electron mobility $\mu = {e\tau}/{m^{*}}$, where $\tau$ is the scattering time and $m^{*}$ is the effective mass (${1}/{m^{*}}=({1/\hbar^{2}}){d^{2}E(k)}/{dk^2}$, where $E(k)$ is the energy dispersion relation \cite{Neaman}.)

$\text{\bf{Li-vacancy Tetragonal LiBO}}_{\bf{2}}{ }(\text{\bf{Livac t-LBO)}}$: Figure \ref{Figure2}b shows the calculated band structure of the Livac t-LBO. At first glance, it appears that there are no obvious modifications of the band structure of the t-LBO polymorph in the presence of Li vacancies. However, a closer look shows the following important changes: 

First, while the minima of the lowest energy dispersion relation $E(k)$ curve of the conduction band are still located at $\Gamma$ and Z $k$-points with the VBM at the former, all of the $E(k)$ curves of the conduction band were shifted up in energy by $\approx 0.5 \text{ eV}$, a considerable amount of energy compared to the thermal energy ($\approx 0.026 \text{ eV}$).

Second, the curvatures of the lowest $E(k)$ curve of the conduction band at the $\Gamma$ and Z $k$-points of the Livac t-LBO (see Figure \ref{Figure2}b) are both smaller than those of the Pri t-LBO (see Figure \ref{Figure2}a), indicating that the Li vacancies reduce the mobility of electrons excited to the conduction band of the m-LBO polymorph.

Third, the CBM of the band structure of the Livac t-LBO is still located at the same $k$-point (i.e., the N $k$-point) as that of the Pri t-LBO, suggesting that the indirect band gap feature is still preserved with the presence of Li vacancies. The Fermi level, however, is shifted down by $\approx 0.2 \text{ eV}$ to the energy levels of the maxima of the highest $E(k)$ of the conduction band at the $\Gamma$ and Z $k$-points. This means that the energy levels introduced by the B vacancies between VBM and the Fermi level manifest the acceptor levels of the B vacancies, and the Livac m-LBO might behave as a p-type semiconductor with a wide band gap of $E_{\text{g}} =7.8 \text{ eV}$, which is 0.5 eV higher than that ($E_{\text{g}} =7.3 \text{ eV}$) of the Pri t-LBO as a result of the uplifting of the conduction band by 0.5 eV as aforementioned. 

Fourth, the curvatures of the highest $E(k)$ of the valence band at its maxima (\textit{i.e.}, at the $\Gamma$, Z, and N $k$-points) are smaller (\textit{i.e.}, the $E(k)$ curve at these $k$-points is flatter for the Bvac t-LBO than for the Pri t-LBO), suggesting that electronic holes (which are excited to the valence band) possess increased effective masses, and hence reduced mobilities, in the presence of Li vacancies in the t-LBO polymorph. 

Finally, let us now consider the application perspective of the Livac t-LBO. In our previous work (see Ref. \cite{Ziemke1stDFT_LiBO2}), we have shown that Li vacancies need to be present in t-LBO crystals for Li-ion transport via a vacancy-mediated diffusion mechanism. Of all the 9 Li-ion diffusion pathways investigated for the Livac t-LBO, the lowest diffusion barrier is $E_{\text{m}} = 0.46 \text{ eV}$ and thus an estimated ionic conductivity is $\sigma_{\text{ion}}=7.0\times10^{-5}\text{ S cm}^{-1}$. Nevertheless, this value of $\sigma_{\text{ion}}$ and the band gap $E_{\text{g}}= 7.8 \text{ eV}$ both show that the Livac t-LBO is not suitable as a potential conformal cathode coating of Li-ion batteries since the coating should be both a good ionic conductor and an electronic conductor.

$\text{\bf{B-vacancy Tetragonal LiBO}}_{\bf{2}}{ }(\text{\bf{Bvac t-LBO)}}$: Figure \ref{Figure2}c shows the band structure of the  Bvac t-LBO. In general, the modification of the band structure induced by B vacancies exhibits similar features to those discussed above for the Livac t-LBO with some minor differences. These features are briefly described as follows:

First, the conduction band of the Bvac t-LBO is shifted up by 0.3 eV, which is less than that (0.5 eV) of the Livac t-LBO while the minima of the lowest $E(k)$ curve of the conduction band are still located at both the $\Gamma$ and Z $k$-points with the CBM located at the $\Gamma$ $k$-point.

Second, the curvatures of the lowest $E(k)$ curve of the conduction band at both the $\Gamma$ and Z $k$-points are smaller than those of the Livac t-LBO and the Pri t-LBO, meaning that the mobility of the electron excited to the conduction band is reduced even further in the presence of B vacancies in the t-LBO polymorph. 

Third, like the case of the Livac t-LBO, the VBM is located at the N $k$-point while the Fermi level is pushed down by 0.14 eV, which is slightly less than that of the Livac t-LBO, meaning that the energy band of the acceptor introduced by B vacancies, which are located between the Fermi level and the VBM (see Figure \ref{Figure2}c), is a bit narrower than that introduced by Li vacancies (see Figure \ref{Figure2}b). Yet, the Bvac t-LBO might still behave as a p-type semiconductor with a wide band gap of $E_{\text{g}}=7.6 \text{ eV}$, which is wider than that ($E_{\text{g}}=7.3 \text{ eV}$) of the Pri t-LBO, but narrower than that ($E_{\text{g}}=7.8 \text{ eV}$) of the Livac t-LBO.

Fourth, the highest $E(k)$ curve of the valence band of the Bvac t-LBO is flatter than that of the Livac t-LBO, meaning that the electron hole of the Bvac t-LBO is heavier with its smaller mobility.

Finally, let us now compare the applicability of the Bvac t-LBO and the Livac t-LBO as a solid electrolyte or a conformal cathode coating of Li-ion batteries. In our previous DFT work (see Ref. \cite{Ziemke1stDFT_LiBO2}), we reported that the Bvac t-LBO has an ionic conductivity ranging from $\sigma_{\text{ion}} = 2.5\times10^{-3} \text{ S cm}^{-1}$ to $\sigma_{\text{ion}} = 7.5 \text{ S cm}^{-1}$, which is in the upper range of values of the ionic conductivity of the currently developed solid electrolytes for all-solid Li-ion batteries and shows that the Bvac t-LBO is promising, in terms of its Li-ion transport, for its applications as both a solid electrolyte and a conformal cathode coating (\textit{i.e.,} a good ionic conductor). However, in terms of the electron/hole transport deduced from our DFT-calculated electronic band structure shown in Figure \ref{Figure2}c, the combined findings of our previous \cite{Ziemke1stDFT_LiBO2} and current works suggest that the Bvac t-LBO would be a potential candidate as a solid electrolyte rather than a conformal cathode coating. 

This is because the Bvac t-LBO possesses a wide band gap ($E_{\text{g}}=7.6 \text{ eV}$) (i.e., a good electronic insulator) which exceeds the high-voltage window (3 to 5 V) of Li-ion batteries. In addition, the acceptor levels are located very close to the top of the valence band that will accept electrons to the sites of B vacancies, which are then negatively charged to facilitate Li-ion transport via a vacancy-mediated diffusion mechanism (\textit{i.e.,} positively-charged Li$^{+}$ ions could be electrostatically driven by the negatively-charged B vacancies to hop into their sites on their diffusion pathways \cite{Ziemke1stDFT_LiBO2}). Yet, a good electronic insulator is not suitable to be a conformal cathode coating since there still is a  certain voltage drop across the coating layer due to a high electronic resistivity. 

$\text{\bf{O-vacancy Tetragonal LiBO}}_{\bf{2}}\text{ \bf{(Ovac t-LBO)}}$: Figure \ref{Figure2}d presents the band structure of the Ovac t-LBO. Compared to Li and B vacancies in the t-LBO polymorph, O vacancies modify the band structure significantly. 

First, the conduction band of the Ovac t-LBO is substantially shifted down in energy by 2.6 eV, instead of shifting up as seen for both the Livac t-LBO (up by 0.5 eV) and the Bvac t-LBO (up by 0.3 eV). 

Second, the shapes of all energy dispersion $E(k)$ curves of the conduction band are completely changed in comparison with those of the Pri t-LBO. The minima of the lowest $E(k)$ curve of the conduction band are moved from the $\Gamma$ and Z $k$-points for the Pri t-LBO to the $\Sigma$ and $\Sigma_{1}$ $k$-points with the CBM located at the $\Sigma$  $k$-point for the Ovac t-LBO. The curvature of the lowest $E(k)$ curve around the $\Sigma$ $k$-point (\textit{i.e.}, at the CBM) for the Ovac t-LBO is the least among the four systems shown in Figure \ref{Figure2}, meaning that the effective masses of electrons in the Ovac t-LBO  are largest and their mobilities are lowest among electrons in each of the four systems of the t-LBO polymorph, whose electronic band structures are shown in Figure \ref{Figure2}. 

Third, the valence band is also substantially modified. The VBM is now moved from the N $k$-point for the Pri t-LBO to the X $k$-point for the Ovac t-LBO. In addition, the VBM is located at the energy level of 1.7 eV below the Fermi level. The highest $E(k)$ curve of the valence band is modified in such a way that it is more wavy than those of the Livac t-LBO and Bvac t-LBO. The energy band gap ($E_{\text{g}}=6.4 \text{ eV}$) is thus less than that for the other three systems. 

Fourth, a remarkable new feature is generated: deep defect energy levels are introduced by oxygen vacancies into the band gap. The bottom of the defect band (DBB) and the top of the defect band (DBT) are respectively located at the N $k$-point (0.7 eV above the VBM) and at the X and P $k$-points (4.7 eV below the CBM). The vacancy sites of O$^{2-}$ anions act as the deep centers to trap electrons. Thus, they were called in the fields of electro- and photo-optics as the color, Forbe, or F-centers \cite{Tilley}. An F center is a type of crystallographic defect that forms when a negative ion is removed from a crystal lattice, leaving behind a vacancy. This vacancy can capture an electron, potentially giving the material a colored appearance. 

Finally, because of electrons being trapped to the deep defect levels of oxygen vacancies, the Fermi energy level is no longer located at the VBM, but rather moves to the top of the defect band (TDB), resulting in a reduced effective indirect band gap, which is as small as $E_{\text{g}}^{\text{eff}}=4.7 \text { eV}$, indicating that the Ovac t-LBO is not as good an electronic insulator as the Bvac t-LBO for use as a solid electrolyte. In addition, our previous work (see Ref. \cite{Ziemke1stDFT_LiBO2}) showed that the ionic conductivity of the Ovac t-LBO is estimated to be $\sigma_{\text{ion}} = 7.6 \times 10^{-8} \text{ S cm}^{-1}$, which is too low to be a good ionic conductor and inhibits the Ovac t-LBO from being either a solid electrolyte or a conformal cathode coating.

\subsection{Effects of Lattice Vacancies on Electronic Band Structure of Monoclinic LiBO$_{2}$ \label{m-LBO}}

\begin{figure}[!ht]
\centering
\includegraphics[width=1.0\linewidth]{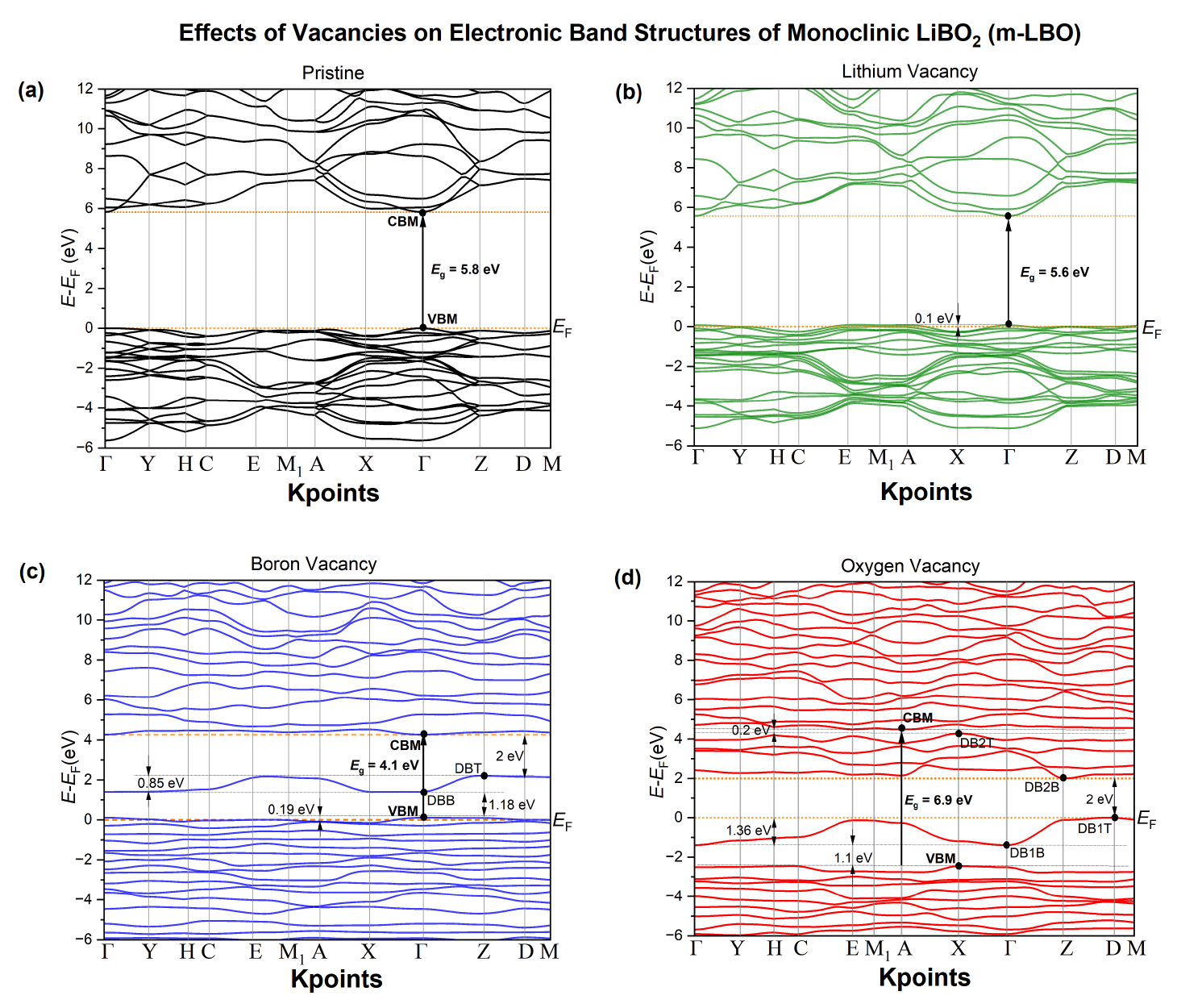}
\caption{\label{Figure3} Calculated electronic band structure of monoclinic LiBO$_{2}$ (m-LBO) for (a) pristine, (b) Li-vacancy, (c) B-vacancy, and (d) O-vacancy supercells. The conduction band minimum, valence band maximum, defect band 1 top, defect band 1 bottom, defect band 2 top, and defect band 2 bottom are abbreviated as CBM, VBM, DB1T, DB1B, DB2T, and DB2B, respectively.}
\end{figure}

Figure \ref{Figure3} shows the calculated electronic band structures for the monoclinic LiBO$_{2}$ polymorph: pristine (Pri m-LBO, Figure \ref{Figure3}a), Li-vacancy (Livac m-LBO, Figure \ref{Figure3}b), B-vacancy (Bvac m-LBO, Figure \ref{Figure3}c), and O-vacancy (Ovac m-LBO, Figure \ref{Figure3}d) supercells.

$\text{\bf{Pristine Monoclinic LiBO}}_{\bf{2}}\text{\bf{ (Pri m-LBO)}}$: As one can see in the Figure \ref{Figure3}a, different from the Pri t-LBO discussed above in Section \ref{t-LBO}, the electronic band structure of the Pri m-LBO exhibits a direct band gap of $E_{\text{g}}=5.8\text{ eV}$. As a typical descriptor of a direct-band-gap intrinsic insulator,  both the conduction band minimum (CBM) and valence band maximum (VBM) are located at the same $\Gamma$ $k$-point, and the Fermi level is aligned with the VBM. In addition, the value of $E_{\text{g}}$ is also less than the value ($E_{\text{g}}=7.6\text{ eV}$) calculated by Basalaev \textit{et al.} \cite{Basalaev2019} using an expensive hybrid functional method. Yet, similar to their calculations, our calculations showed that the value of $E_{\text{g}}$ of the Pri m-LBO is smaller than that of the Pri t-LBO, even though both were underestimated due to the limitation of the GGA approximation we used in this work. Finally, it is suggested that while both the Pri m-LBO and the Pri t-LBO are good electronic insulators, the latter apparently insulates electrons even better. 

$\text{\bf{Li-vacancy Monoclinic LiBO}}_{\bf{2}}\text{\bf{ (Livac m-LBO)}}$: The electronic band structure of the Livac m-LBO is slightly modified from that of the Pri m-LBO in the presence of Li lattice vacancies, as shown in Figure \ref{Figure3}b.

First, the material retains a direct band gap at the $\Gamma$ $k$-point, but the band gap energy is slightly reduced to $E_{\text{g}}=5.6\text{ eV}$, leading to an increased electron concentration, $n_{\text{e}}(T)$, at temperature $T$ since $n_{\text{e}}(T)\propto e^{-E_{\text{g}}/k_{\text{B}}T}$ \cite{Neaman}. Specifically, it is shown that at the room temperature ($T = 300 {\text{ K}} =  0.026 \text{ eV}$), ${n_{\text{e}}^{\text{Livac n-LBO}}}/{n_{\text{e}}^{\text{Pri m-LBO}}} \approx e^{-5.6 \text{ eV}/0.026 {\text{ eV}}} / e^{-5.8 \text{ eV}/0.026 {\text{ eV}}} \approx 2.2 \times 10^{3}$. This estimated ratio suggests that $n_{\text{e}}$ of the m-LBO polymorph might be increased by 1000 times at 300 K in the presence of Li vacancies. 

Second, it is noted that the lowest energy dispersion relation curve $E(k)$ of the conduction band near the CBM becomes  flatter, indicating a slightly reduced curvature of $E(k)$ vs. $k$ at the $\Gamma$ $k$-point and consequently an increased electron effective mass and hence a reduced electron mobility. 

Third, the Fermi level shifts down by 0.2 eV below the VBM, and the CBM lies at a slightly lower energy compared to the pristine structure. The defect energy states introduced by Li vacancies are shallowly positioned between the VBM and Fermi level, indicating that the Li vacancies act as electron acceptors and therefore the Livac m-LBO resembles a wide band-gap p-type semiconductor.

Finally, to assess the Livac m-LBO's applicability, we recall its Li-ion transport properties from Ref. \cite{Ziemke1stDFT_LiBO2}: its ionic conductivity is $\sigma_{\text{ion}} = 7.0 \times 10^{-13} \text{ S cm}^{-1}$, seven orders of magnitude lower than the $\sigma_{\text{ion}} = 7.0 \times 10^{-5} \text{ S cm}^{-1}$ of the Livac t-LBO. While both are lower than the ionic conductivity of typical Li-ion battery liquid electrolytes ($\sigma_{\text{ion}} \sim 10^{-2} \text{ to 10 S cm}^{-1}$ \cite{Zhang2018}), our comparison suggests that the Livac t-LBO is a better choice for a solid electrolyte due to its superior ionic conduction and electron insulation.

$\text{\bf{B-vacancy Monoclinic LiBO}}_{\bf{2}}\text{ \bf{(Bvac m-LBO)}}$: Unlike Li vacancies discussed above, B vacancies in the m-LBO lattice modify its electronic band structure substantially. 

First, Figure \ref{Figure3}c shows that the bottom of the conduction band of the Bvac m-LBO shifts down in energy by $\sim$ 2 eV relative to that of the Pri m-LBO while the VBM is still located at the center of the first Brillouin zone (\textit{i.e.,} the $\Gamma$ $k$-point). Nevertheless, the curvature of the lowest energy dispersion relation $E(k)$ curve of the conduction band at the CBM (\textit{i.e.,} at the $\Gamma$ $k$-point) is substantially reduced in comparison to those of both the Pri m-LBO and the Livac m-LBO, revealing a larger effective mass, and hence a smaller electron mobility, than those of both the Pri m-LBO and the Livac m-LBO. 

Second, the VBM is still at the $\Gamma$ $k$-point, but 0.2 eV above the Fermi level, indicating a direct band gap of $E_{\text{g}}\approx4\text{ eV}$, significantly smaller than those of both the Pri m-LBO and Livac m-BO. Like the Livac m-LBO, the empty energy levels between the Fermi level and the VBM, which are introduced into the band gap by B vacancies, manifest shallow acceptor levels, typically occurring in the electronic band structure of a p-type wide band-gap semiconductor. 

Third, there is a $E(k)$ curve in the middle region of the band gap. This new curve is attributed to the presence of B-vacancy-induced deep levels that collectively form a defect band of 0.85 eV in width, whose bottom-most level (abbreviated as DBB) is at the $\Gamma$ $k$-point and is located at the energy level 1.18 eV above the VBM level and whose top-most level (abbreviated as DBT) is at the Z $k$-point and is located 2 eV below the CBM level. Unlike shallow levels, deep levels are located significantly far from the conduction or valence band edges and have highly localized wave functions, meaning the trapped charge carriers (in this case, the carriers are electronic holes rather than electrons as in the Ovac m-LBO discussed below) are tightly bound to defect sites (\textit{i.e.}, B vacancies). 
At zero temperature, these deep levels are unoccupied with electrons (\textit{i.e.,} empty states), suggesting that B vacancies may act as deep centers that trap electronic holes instead of electrons (\textit{i.e.,} hole trapping centers) as shown in Figure \ref{Figure3}c. 
However, at a finite temperature, such as the room temperature at which electric devices like Li-ion batteries normally operate, these deep centers can act differently, depending on the amount of energy, $\Delta E$, electrically (\textit{i.e.}, via a bias dc electric field) or optically (\textit{i.e.}, via excitation photons) imparted to electrons. If $\Delta E < 1.18 \text{ 
eV}$ B vacancies behave as hole-trapping centers. 
In contrast, if $1.8\text{ eV}\leq\Delta E < 2\text{ eV}$ they act as electron-trapping centers, negatively-charged B vacancies thus can attract positively-charged Li$^{+}$ ions that are on their pathways to transport toward the vacancies, facilitating Li-ion diffusion via a vacancy-mediated mechanism as reported in our previous work \cite{Ziemke1stDFT_LiBO2}. 
However, if $\Delta E \geq 2 \text{ eV}$, they become recombination centers for both electrons and electronic holes, playing the role of stepping stones for charge carriers (e.g., Li$^{+}$ ions, electrons, and electronic holes) to jump between sites in the Bvac m-LiBO lattice, facilitating the carriers to conduct an electric current upon the application of a sufficiently strong bias voltage (\textit{e.g.,} the voltage across a LiBO$_{2}$ solid electrolyte  sandwiched between the cathode and anode of Li-ion battery cells operating at voltages higher than 3 V), or facilitating a photocurrent upon the absorption of UV-Vis photons.
For $\Delta E \geq 2 \text{ eV}$, both electrons and electronic holes can participate in the electric conduction; however, electronic holes are the major carriers and electrons are the minor carriers. This is because the number of electronic holes per unit volume (\textit{i.e.,} the concentration), 
$n_{\text{h}}(T)$ $\propto$  $e^{-(E_{\text{DBB}}-E_{\text{VBM}})/{k_{\text{B}}T}}$ 
= $e^{-{1.18\text{ eV}}/{k_{\text{B}}T}}$,
is much greater than that of electrons, 
$n_{\text{e}}(T) \propto e^{-\frac{E_{\text{CBM}} - E_{\text{DBT}}}{k_{\text{B}}T}}$
= $e^{-2 \text{ eV}/k_{\text{B}}T}$. 
For example, at room temperature ($T = 300 {\text{ K}} =  0.026 \text{ eV}$), $n_{\text{h}}/n_{\text{e}} \approx 4.6 \times 10^{13}$, meaning that the concentration of electronic holes is 13 orders of magnitude greater than that of electrons in the presence of B vacancies in the lattice of the m-LBO polymorph. 

Finally, from the application points of view, in order to select the Bvac m-LBO as a solid electrolyte (a good electronic insulator) or/and a conformal cathode coating (a good electronic conductor), one must first consider the electron transport with an effective band gap for the electron excitation at a finite temperature. This effective band gap would be estimated as $E_{\text{g}}^{\text{eff}} =E_{\text{CMB}}-E_{\text{DBT}} = 2 \text{ eV}$ (see Figure \ref{Figure3}c),  which is much smaller than those of the Pri m-LBO ($E_{\text{g}} = 5.8 \text{ eV}$) and the Livac m-LBO ($E_{\text{g}} = 5.8 \text{ eV}$). Effectively, the excitation of electrons from the "stepping stone" energy level of the defect band top (DBT) at the Z $k$-point to the CBM at the $\Gamma$ $k$-point manifests an indirect-band scenario. In addition, as reported in our previous work
\cite{Ziemke1stDFT_LiBO2}, among the defective systems of the m-LBO polymorph considered in this section (Section \ref{m-LBO}), the Bvac m-LBO possesses the highest Li-ionic conductivity ($\sigma_{\text{ion}}=2.8\times 10^{-2} \text{ S cm}^{-1}$), which is about the lower bound of the ionic conductivity values of typical commercialized liquid electrolytes of Li-ion batteries ($\sigma_{\text{ion}} \sim 10^{-2} \text{ to 10 S cm}^{-1}$ \cite{Zhang2018}). Overall, it is suggested that the Bvac m-LBO might be a promising candidate for the conformal cathode coating of Li-ion batteries (a good Li-ionic conductor and a good electronic conductor as well), not as a solid electrolyte (a good Li-ionic conductor and a good electronic insulator). 

$\text{\bf{O-vacancy Monoclinic LiBO}}_{\bf{2}}\text{ \bf{(Ovac m-LBO)}}$: Figure \ref{Figure3}d shows that O vacancies substantially modify the electronic band structure of the m-LBO polymorph. 

First, similar to the Ovac t-LBO (see Figure \ref{Figure2}d), the VBM of the band structure of the Ovac m-LBO is moved from the $\Gamma$ $k$-point to the X $k$-point. In contrast, this movement does not happen for both the Livac m-LBO (see Figure \ref{Figure3}b) and Bvac m-LBO (see Figure \ref{Figure3}c) for which the VBM was still located at the same $\Gamma$ $k$-point as observed for the Pri m-LBO (see Figure \ref{Figure3}a).

Second, like the Ovac t-LBO, the energy level of the VBM of the Ovac m-LBO is not at the Fermi level, but about 2.46 eV below it. This is in contrast to the Bvac m-LBO (see Figure \ref{Figure3}c) for which the Fermi level is aligned with the VBM. 

Third, similar to the Ovac t-LBO (see Figure \ref{Figure2}d) and the Bvac m-LBO (see Figure \ref{Figure3}c), there is also a curve of the energy dispersion relation $E(k)$ in the middle of the band gap of the Ovac m-LBO. This curve is attributed to one of the defect bands  (a deep-level defect band or defect band 1, DB1), which is introduced by O vacancies and spans an energy range of 1.36 eV from the defect band bottom (DB1B) (which is located at the $\Gamma$ $k$-point and 1.1 eV above the VBM) to the defect band top (DB1T) (which is 2 eV below the bottom of another defect band introduced by oxygen vacancies). Similar to the Ovac t-LBO (see Figure \ref{Figure2}d), the Fermi level is located at the DB1T rather than at the VBM, meaning that these deep levels also act as the electron trapping centers or F centers. However, unlike the band structure of the Ovac t-LBO, there is another defect band, the defect band 2 (DB2), introduced into the band gap of the band structure of the Ovac m-LBO, just 0.2 eV below the CBM. The DB2 acts as an effective conduction band while the DB1 acts as an effective valence band. Thus, the band gap is effectively reduced to $E_{\text{g}}^{\text{eff}} = 2 \text{ eV}$ from the nominal band gap of $E_{\text{g}} = 6.9 \text{ eV}$. As a result, similar to the Bvac m-LBO, the Ovac m-LBO could be considered as a good electronic conductor for its use as a conformal cathode coating with the condition that its ionic conductivity must be high enough. However, our previous work reported that the Ovac m-LBO is a poor Li-ion conductor with the critically low ionic conductivity of $\sigma_{\text{ion}}=6.1\times 10^{-6} \text{ S cm}^{-1}$ (Ref. \cite{Ziemke1stDFT_LiBO2}) compared to  the ionic conductivity values of typical commercialized liquid electrolytes of Li-ion batteries ($\sigma_{\text{ion}} \sim 10^{-2} \text{ to 10 S cm}^{-1}$ \cite{Zhang2018}). Overall, the Ovac m-LBO should be neither a solid electrolyte nor a conformal cathode coating of Li-ion batteries.

In summary, the combined outcomes of our current work and previous one (see Ref. \cite{Ziemke1stDFT_LiBO2}) suggest that the Bvac t-LBO might be a promising candidate for the solid electrolyte of Li-ion batteries, while its Bvac m-LBO counterpart has potential as a conformal cathode coating of Li-ion batteries.

\section{Conclusions \label{conclusions}}

In conclusion, our studies highlight the potential of B vacancies in the LiBO$_{2}$ material for enhancing its performance in lithium-ion battery applications. By improving the electronic insulation of the t-LBO polymorph and the electronic conduction of the m-LBO polymorph while boosting the ionic conductivity of both polymorphs, B lattice vacancy engineering emerges as a promising strategy for optimizing the properties of LiBO$_{2}$ as either a solid electrolyte (t-LBO) or a conformal cathode coating (m-LBO) for its use in Li-ion batteries. In reality, as aforementioned in the Introduction, m-LBO rather than t-LBO is stable at ambient temperature and pressure, it is therefore suggested from our works that the LiBO$_{2}$ material in the presence of B vacancies would be considered to be a potential conformal cathode coating in the form of crystalline films.  It is noted that cathode coatings were reported to be partially or fully amorphous, though their structures largely depend on the specific synthesis methods and conditions. In some cases, these methods may lead to coatings with nanocrystallite morphologies. Small-particle-on-large-particle coating methods reported by Chen and Dahn \cite{Chen2002} and Zhao \textit{et al.} \cite{Zhao2012} are examples of the formation of crystalline coatings on the cathode surface. Nevertheless, this work has only focused on the crystalline polymorphs of the LiBO$_{2}$ coating rather than its amorphous ones, as the former are the simplest to study using atomistic modeling techniques and serve as a benchmark for assessing coating performance in more complex thin films and their interface with the electrodes \cite{Xu2015}. Overall, our combined DFT findings underscore the importance of lattice vacancy manipulation, particularly through techniques like thermal neutron irradiation, in advancing LiBO$_{2}$ material as a highly functional material for next-generation energy storage technologies.

\section*{Conflicts of Interest} 
The authors declare no conflict of interest.

\section*{Author Contributions}

The contributions of the author to the current work are as follows:

Conceptualization: H. M. N., C.W., T. W. H, Y. X, and J. G;

Methodology: H M. N., C. Z., N.N., and C. W;

Manuscript Writing: H.M.N., C.W., and C.Z.;

Manuscript Proofreading and Reviewing: all authors

Equal Contributions: H. M. N and C. Z..

\section*{Acknowledgments}
 This work was funded in part by the National Science Foundation Grant No. IIP-2044726, the University of Missouri Materials Science and Engineering Institute (MUMSEI) Grant No. CD002339, and the University of Missouri Research Reactor (MURR). The computation for this work was performed on the high-performance computing infrastructure provided by Research Support Services at the University of Missouri, Columbia, MO. DOI: https://doi.org/10.32469/10355/97710. The careful manuscript proofreading and valuable comments of Dr. Bikash Saha (MURR) to improve the quality of this work are greatly appreciated.



\end{document}